\documentclass[journal]{IEEEtran}
\usepackage{cite}

\ifCLASSINFOpdf
  \usepackage[pdftex]{graphicx}
\else
\fi
\usepackage{amsmath}
\usepackage{array}

\usepackage[utf8]{inputenc}

\begin{document}
%
\title{Cavity Simulator for European Spallation Source}
%
%
%

\author{Maciej~Grzegrzółka,
        Krzysztof~Czuba,~\IEEEmembership{Member,~IEEE,}
	  Mateusz~Lipiński,
        and~Igor~Rutkowski,
\thanks{Maciej Grzegrzółka, Krzysztof Czuba, Mateusz Lipiński and Igor Rutkowski are with Warsaw University of Technology
(WUT), Warsaw, Poland..}
\thanks{Manuscript received June 24, 2018}}

\maketitle

\begin{abstract}
European Spallation Source will be the brightest neutron source in the world. It is being built in Lund, Sweden. Over 120 superconducting cavities will be installed in the facility, each regulated by an individual LLRF control system. To reduce the risk associated with testing the systems on real cavities a Cavity Simulator was designed. It reproduces the behaviour of superconducting cavities used in the medium and high beta sections of ESS's Linac. The high power RF amplifier and piezo actuators parameters are also simulated.

Based on the RF drive and piezo control signals the Cavity Simulator generates the RF signals acquired by the inputs of the LLRF control system. This is used to close the LLRF feedback loop in real time. The RF front end of the Cavity Simulator consists of vector modulators, down-converting circuits, and a set of fast data converters. The cavity response simulation is performed in a high speed FPGA logic by a dedicated firmware, that was optimized to minimize the processing time. The device also generates clock, LO, and the 704.42 MHz reference signals to allow for system tests outside of the accelerator environment. 

In this paper the design of the Cavity Simulator, description of the algorithms used in its firmware, and measurement results of the device are presented.

\end{abstract}


%
\IEEEpeerreviewmaketitle

\section{Introduction}

\IEEEPARstart{M}{edium} and high beta sections of European Spallation Source linac will consist of 120 superconducting elliptical cavities operating at 704.42 MHz \cite{linac}. Each of these cavities requires an individual low level radio frequency (LLRF) control system. Lund University and ESS design the architecture of those systems \cite{LLRF}. They are partnered by Polish Electronics Group (PEG), which develops three of the system's component and will also assemble, test, and install the systems in the ESS facility \cite{PEG}. 

Testing the LLRF control systems with the real cavities is very risky and in can result in an expensive damage. To mitigate this risk, PEG develops the Cavity Simulator which is a device that reproduces the behavior of the real superconducting cavity driven by a high power amplifier. It will be used to test all 704.42 MHz LLRF control system units before they are commisioned with real cavities. This device can also be used for the development of LLRF system's algorithms and firmware. 

\begin{figure}[!htb]
   \centering
   \includegraphics*[width=70mm]{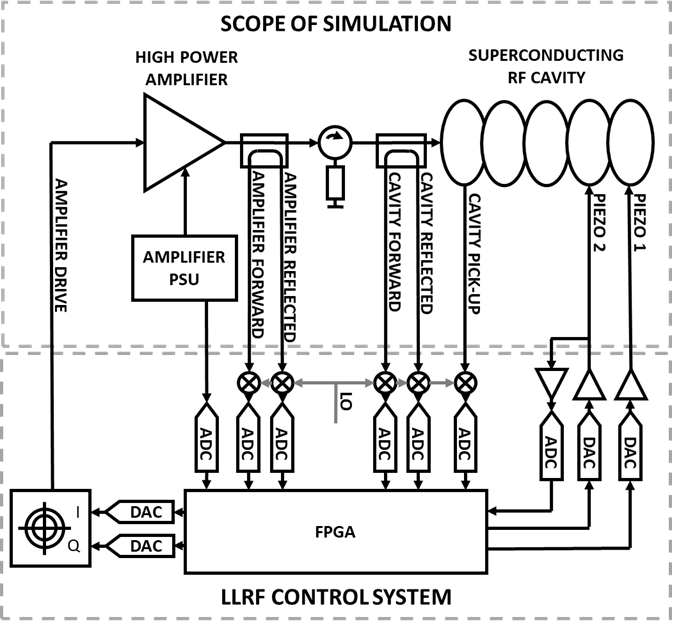}
   \caption{Scope of the simulation.}
   \label{scope}
\end{figure}

Among others, the device simulates following phenomena: cavity dynamics, cavity detuning, piezo compensation, Lorentz force detuning, beam loading, microphonics, amplifier nonlinearity, and amplifiers PSU modulator influence. Fig. \ref{scope}. presents the simplified schematic of the simulated devices and their connection to the LLRF control system.

The simulation model is implemented inside a high-performance FPGA logic, which is combined with a set of data converters and a dedicated analog front-end connecting the Cavity Simulator and the LLRF control system. The simulation parameters can be set on-line through an Ethernet network.

The 704.42 MHz superconducting cavities installed in the Medium and High Beta sections of the ESS Linac are described in  \cite{cavity}. Table \ref{cavtab} lists the essential parameters of those cavities.

Several cavity simulators have been developed by, e.g., DESY, LBNL, and KEK \cite{DESY,LBNL,KEK}. They focus mostly on verifying the algorithms used in the LLRF control systems. The Cavity Simulator presented in this paper allows testing the complete LLRF control system including analog circuits like, e.g., Piezo Driver and ADC frontend. 

This paper has been divided into three parts. The first part presents the Cavity Simulator's hardware design. The firmware and simulation model description follow it. In the final part, the measurements results are shown.

\begin{table}[!htb]
\caption{Basic Parameters of the cavites used in the Medium and High Beta sections of the ESS Linac.}
\begin{tabular}{ccc}
\hline
 Parameter & Medium Beta & High Beta \\
\hline
Nominal Accelerating Gradient (MV/m) & $<$ 16.7 & $<$ 19.9 \\

Number of Cells & 6 & 5 \\

$Q_0$ & $>$ 5e9 & $>$ 5e9 \\

$Q_{ext}$ & 7.5e5 & 7.6e5 \\

Maximum r/Q ($\Omega$) & 394 & 477 \\

$\pi$ and 4$\pi$/5 modes seperation(MHz) & 0.54 & 1.2 \\
\hline
\end{tabular}
 \label{cavtab}
\end{table}

\begin{figure*}[!htb]
  \centering
   \includegraphics*[width=160mm]{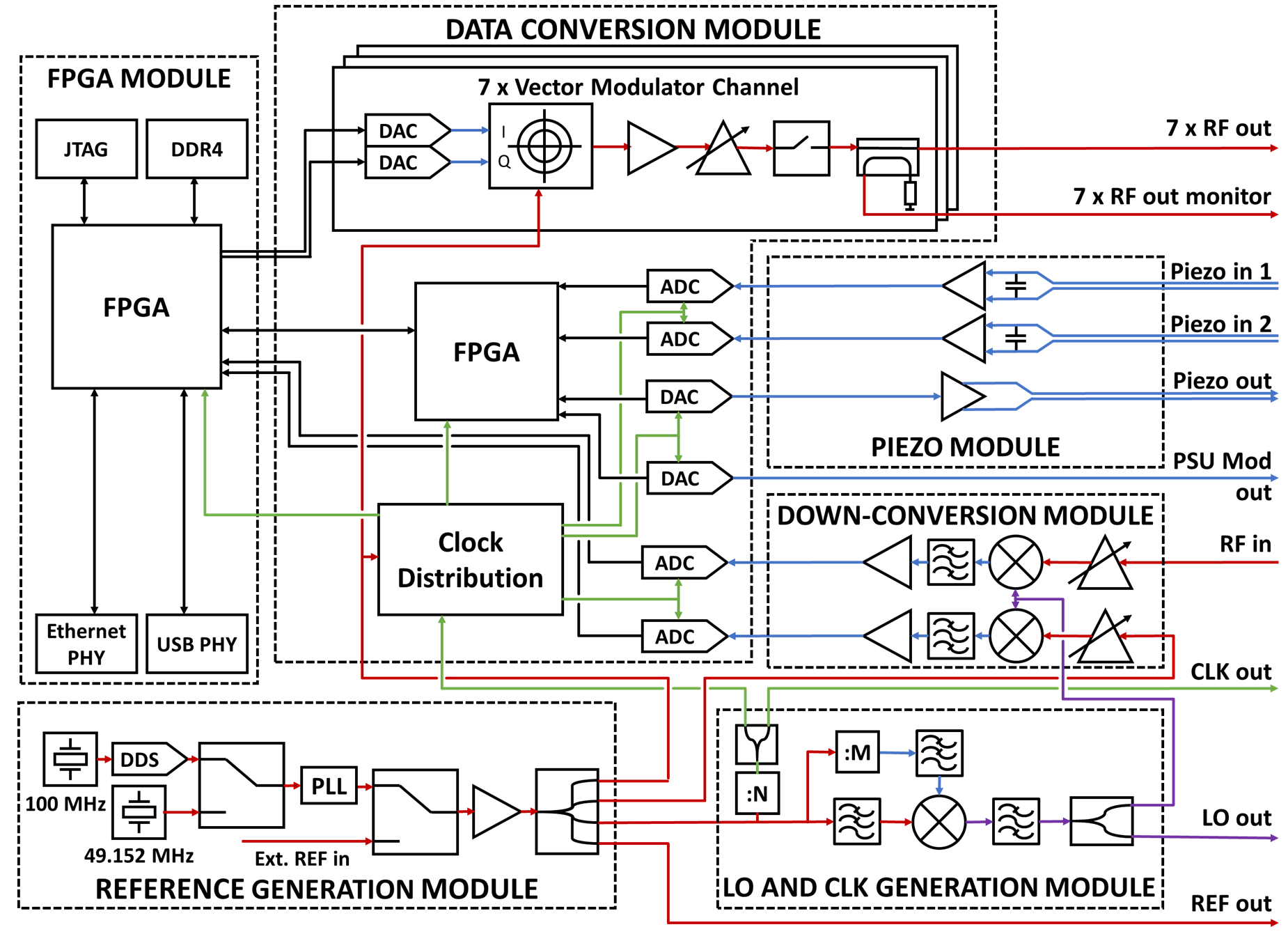}
   \hfill
   \caption{The block diagram of Cavity Simulator's hardware.}
   \label{hardware}
\end{figure*}

\section{Hardware}

The primary purpose of the Cavity Simulator`s hardware is to digitize and generate RF as well as baseband signals. It is also used to produce the reference, LO, and clock signals for the LLRF control system. Additionally, the USB and Ethernet communication interfaces are provided. The device is combined of seven different modules. Fig. \ref{hardware} presents the block diagram showing the structure of the hardware (without the Power Supply Module).

\subsection{FPGA Module}

A vast number of high-performance FPGA modules is available on the market, so to reduce the design time and cost it was decided to use an off-the-shelf component. Xilinx KCU105 board was selected due to a high number of IO pins available and integrated USB and Ethernet PHYs.

\subsection{Data Conversion Module}

The Cavity Simulator requires many different data converters. All of them are located on the Data Conversion Module, which communicates with FPGA module through two FMC connectors. 

In total four analog signals are digitized: RF drive, RF reference, and two piezo drives. The RF signals are typically detected in one of 3 ways:
\begin{itemize}
\item direct sampling,
\item down-conversion,
\item analog IQ demodulation.
\end{itemize}

It was decided that the best method to adopt for this project was using the down-conversion scheme. It is typically the best in terms of performance but requires additional LO generation circuit, which would have to be integrated into the Cavity Simulator nevertheless. The down-converter circuit is designed as a separate module and two 16-bit 250 MSPS ADCs are used to sample the intermediate frequency (IF) output signal.

The piezo drive signals have a bandwidth in a range of tens of kHz. They can be sampled directly without any frequency conversion circuit, but due to a high voltage, analog front-end with additional protection is required. Two 18-bit 5 MSPS ADCs digitize the piezo drive signals.

The Cavity Simulator generates eight analog signals: 6 RF, piezo sensor and a amplifier PSU modulator output. Last two signals do not require any sophisticated front-end and can be generated directly by the DACs. The RF signals are generated using vector modulator circuits. For the best performance, it requires precise length matching of I and Q signal paths, which cannot be achieved using a cable connection. Therefore the vector modulator together with DACs is integrated into the Data Conversion Module. Additional 7th RF output was added to the design. It can be used for the Cavity Simulator development and testing.

The Data Conversion Module is also equipped with a low-cost FPGA. It is used to configure all ICs used in the module and to serialize the data from data converters operating on piezo and PSU modulator signals.

\subsection{Down-conversion Module}

This module lowers the frequency of the input RF signals to a range suitable for the ADCs. It integrates two down-conversion channels, which are based on an active mixer. Each channel has individually controlled gain. Additionally, the mixer's LO signal power level can be controlled and optimized for the best noise performance.

\subsection{Piezo Module}

The amplitude of the piezo drive signals coming from the LLRF Control System can reach 200V. Such voltage can damage the Cavity Simulator's electronics. The Piezo Module lowers the voltage of the piezo signals 100 times and provides additional protection circuit that should withstand the voltages up to 1kV. 

The piezo module also simulates Noliac NAC 2022 H30 piezo used in medium and high beta cryomodules. At cryogenic temperatures, the capacitance of this actuator is around 2.2 $\mu$F.

The Piezo Module also integrates one output channel simulating the piezo working as a detuning sensor. This output is also protected against the high voltage because the Piezo driver used in the ESS' LLRF control systems shares the same connector for both piezo actuator and sensor modes of operation. 

\subsection{Reference Generation Module}

Reference Generation Module distributes the 704.42 MHz RF reference signal, which can be provided externally or internally produced by this module. The generation circuit utilizes a PLL with an integrated VCO. The PLL circuit can be driven by either 49.152 MHz crystal based generator or a DDS operating with 100 MHz clock. Such configuration allows the module to operate in one of 3 modes:
\begin{itemize}
\item generating 704.51 MHz directly from crystal generator signal,
\item generating 704.42 MHz from crystal generator signal using fractional mode of the PLL,
\item generating 704.42 MHz from 22.013 MHz signal produced by DDS.
\end{itemize}

The first option offers the best phase noise performance (78.3 fs jitter in the 10 Hz to 10 MHz integration bandwidth) but has a frequency offset of around 100 kHz from the nominal ESS' RF reference frequency. It is negligible for most applications (under the assumption, that LLRF system uses the same reference signal). In a case where exactly 704.42 MHz frequency is required the second and third options can be used.

\subsection{LO Generation Module}

This module is responsible for generating the LO and Clock signals. The generation scheme is based on direct analog synthesis. The clock frequency is set to 117.4 MHz, and two LO frequencies can be produced: 729.58 and 736.44 MHz. A more detailed description of this module can be found in: \cite{LO}.

\subsection{Power Supply Module}

The Power Supply module powers all other boards used in Cavity Simulator with +12V. Its primary function is conditioning the voltage coming from an off-the-shelf AC-DC converter. Additionally, this module supervises the powering sequence, monitors the power consumption and adjusts the cooling fans speed. 

\subsection{Mechanical Design}

\begin{figure}[!htb]
  \centering
   \includegraphics*[width=70mm]{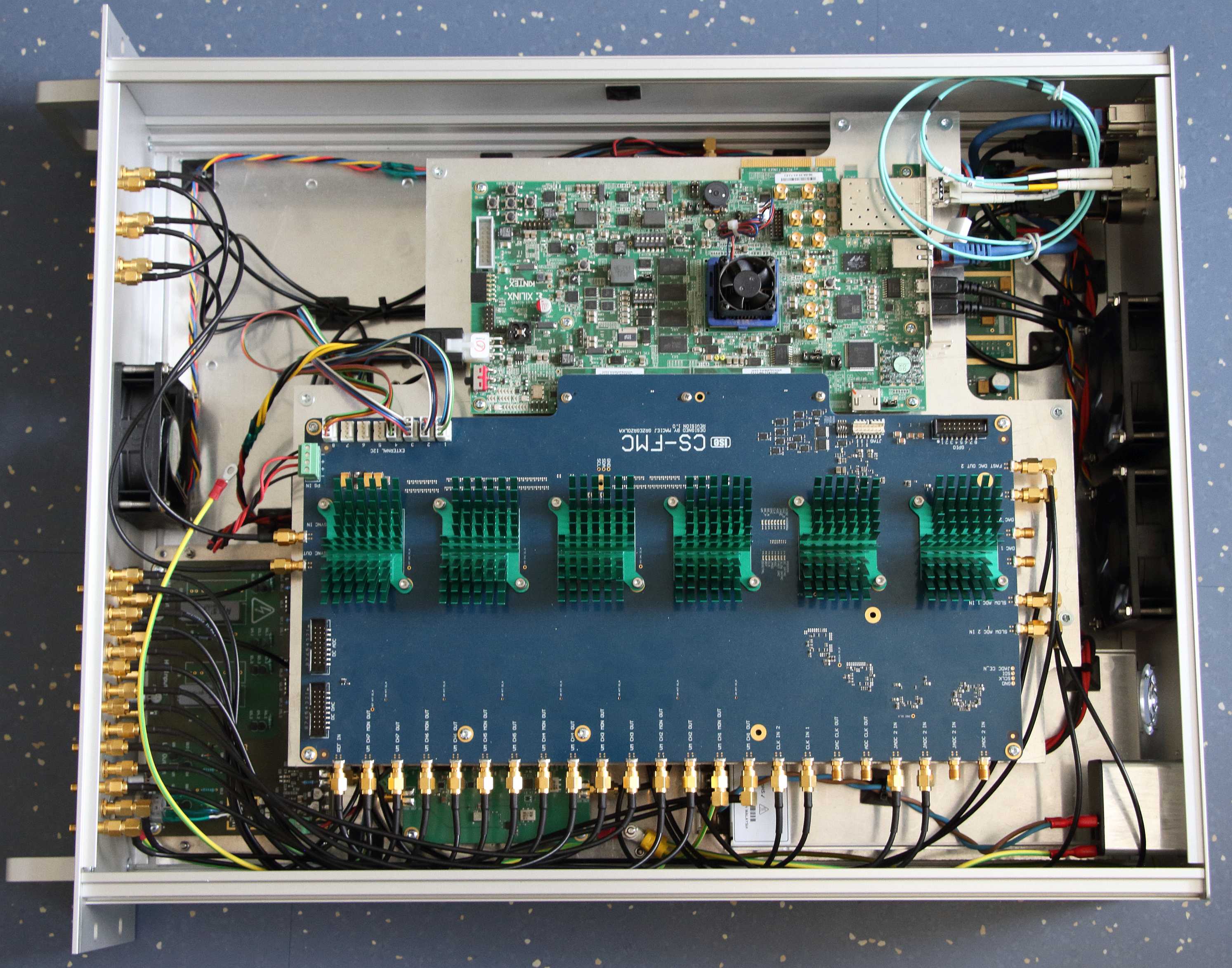}
   \caption{Interior of the Cavity Simulator.}
   \label{foto}
\end{figure}

The Cavity Simulator is integrated into custom 19” rack enclosure. The boards are placed on two floors. On the top FPGA and Data Conversion modules are located. The other modules, including the AC-DC converter, are installed on the lower level. Fig. \ref{foto} shows the interior of the Cavity Simulator.

\subsection{Hardware tests}

\begin{figure}[!htb]
  \centering
   \includegraphics*[width=70mm]{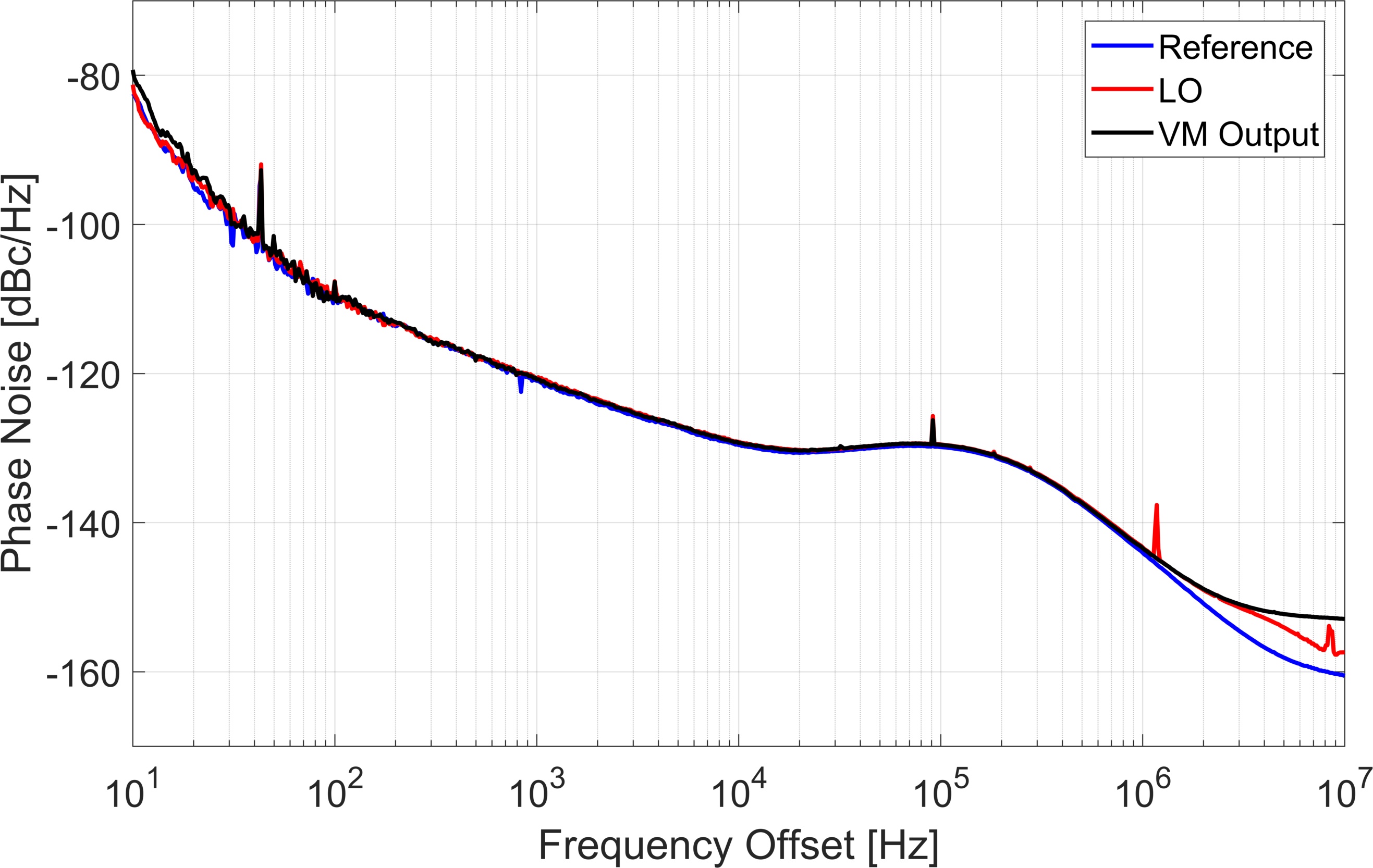}
   \caption{Phase noise spectra of the RF reference, LO and RF output signals.}
   \label{Pnoise}
\end{figure}

The hardware design was tested, and its proper operation was confirmed. The phase noise spectra of the RF reference, LO, and one of the RF outputs signals was measured (see Fig. \ref{Pnoise}). No major interference or additional noise sources were found. The spurs at 42 Hz and 90 kHz offsets are probably originating from the Agilent E5052B Signal Source Analyzer that was used for the measurement. The spur visible in the LO signal's phase noise at 1.1 MHz offset is caused by the DC-DC converter located on the LO Generation Module.

\section{Firmware}
\label{Firmware}

\begin{figure*}[!htb]
   \centering
   \includegraphics*[width=160mm]{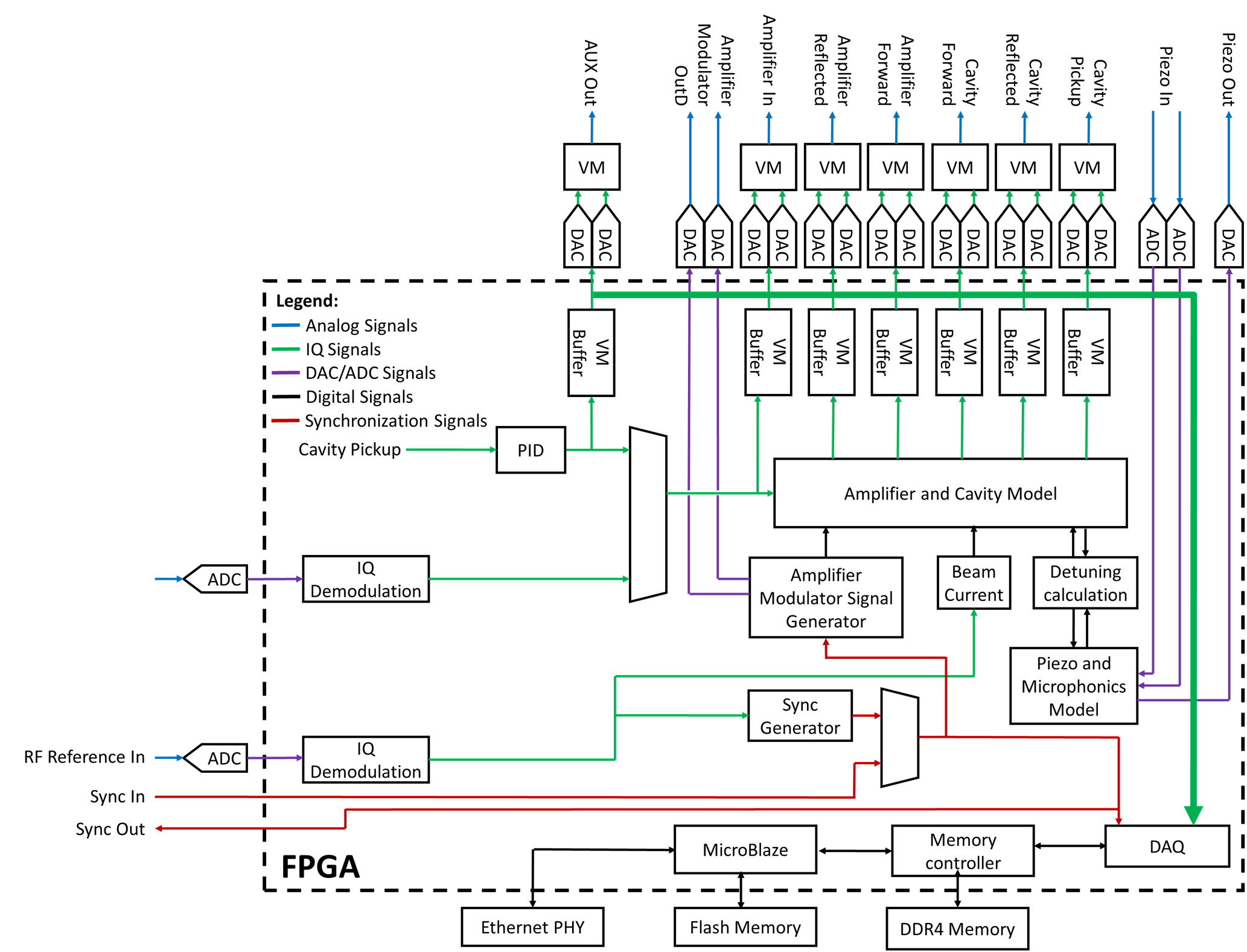}
   \hfill
   \caption{The block diagram of Cavity Simulator's firmware.}
   \label{firmware}
\end{figure*}

Two FPGAs are used in the design. The firmware for the one located on the Data Conversion Module integrates only basic functions like, e.g., SPI and I2C controllers and it will not be described in detail. The firmware for the FPGA Processing Module (see Fig. \ref{firmware}) is much more complicated, and it is responsible for:
\begin{itemize}
\item communication through USB and Ethernet interfaces,
\item gathering data from ADCs,
\item sending data to DACs,
\item data acquisition,
\item digital signal processing for cavity simulation.
\end{itemize}

A Xilinx MicroBlaze softcore microprocessor is used for communication with external systems. It is running a custom software responsible for the interpretation of the control commands sent to the device, setting the parameters of the simulation, and readout of the recorded data stored in the DDR4 memory.

A trigger signal is required to synchronize the Cavity Simulation with the LLRF control system. It can be generated locally, or an external signal can be used. It is fed to the amplifier and cavity model and data acquisition blocks. 

An arbitrary waveform generator driving the additional, 7th RF output is implemented in the firmware. This tool can be used to verify the proper operation of the Cavity Simulator and will simplify device testing. 

The data gathered from the ADCs sampling the RF drive and RF reference is first demodulated using the non-IQ scheme \cite{noniq}. The signals are then processed as a complex vector, allowing the model to operate in the baseband.

One of the critical aspects of the firmware is the processing time. The expected delay introduced mostly by the cabling between the LLRF control system and the cavity is around 400 ns. The total processing time of the whole Cavity Simulator, including analog circuits and data converters latency, shall match this value. The estimated latency introduced by the analog circuits and data converters is around 250 ns. This means that the response of the model must be calculated in less than 18 clock cycles.

\begin{figure*}[!htb]
   \centering
   \includegraphics*[width=160mm]{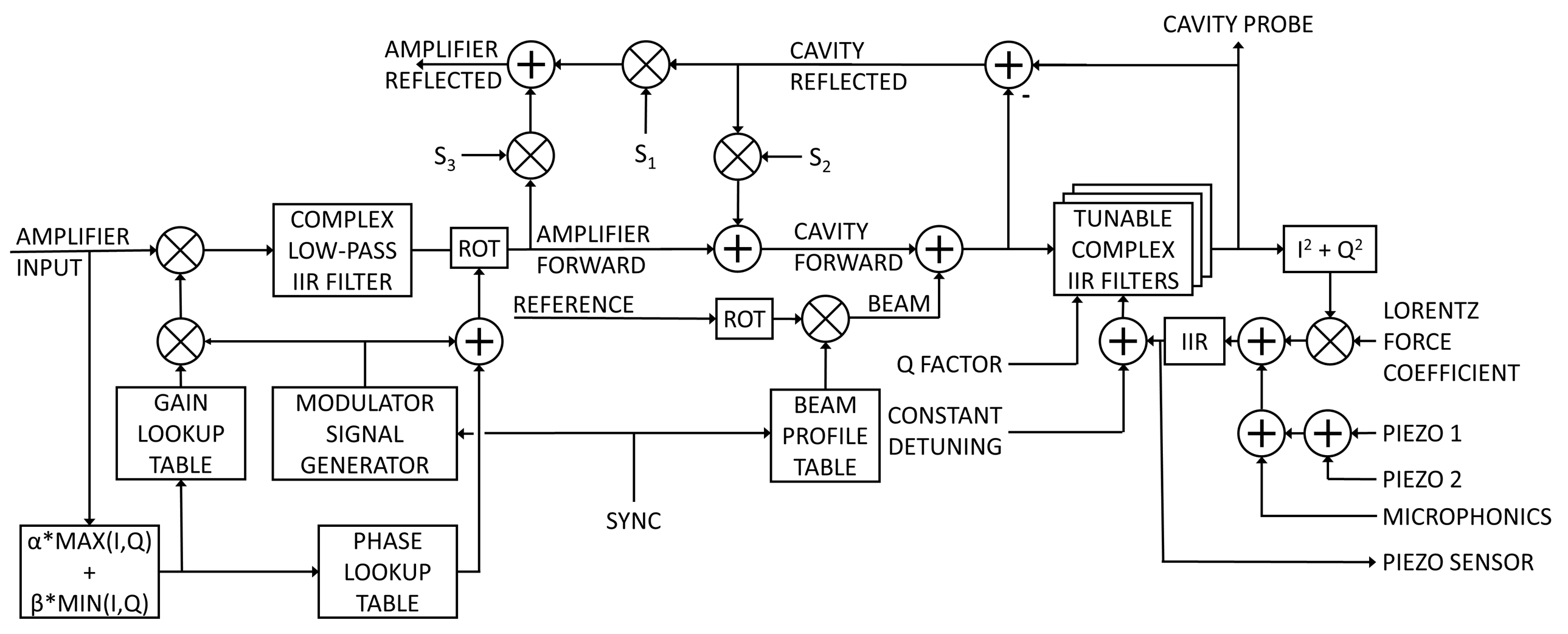}
   \hfill
   \caption{The block diagram presenting the simulation model.}
   \label{model}
\end{figure*}

The primary function of the firmware is to simulate the behavior of the RF cavity, high-power amplifier, and a connection between them through a circulator and a waveguide. The block diagram presenting the model of those elements is shown in Fig. \ref{model}.

\subsection{Cavity Model}

The RF cavity is modeled as a parallel LCR circuit \cite{Schilcher}, which impedance can be expressed using the following relation:

\begin{equation}
Z(\omega) = \frac{1}{\frac{1}{R_L} + \frac{1}{j\omega L} + j\omega C} = \frac{R_L}{1 - jR_L(\frac{1}{\omega L} - \omega C)}
\end{equation}

The L and C can be replaced with:

\begin{equation}
L = \frac{R_L}{Q_L\omega_0} 
\end{equation}

\begin{equation}
C = \frac{Q_L}{\omega_0 R_L}
\end{equation}

where: $Q_L$ - is Q factor of the loaded Cavity and $\omega_0$ is angular frequency of the cavity resonance.  

\begin{equation}
Z(\omega) = \frac{R_L}{1 - jQ_L(\frac{\omega_0}{\omega} - \frac{\omega}{\omega_0})} 
\end{equation}

The signals in the firmware will be operated in the baseband therefore it will be convenient to express the $\omega$ as $\omega_0 + \Delta \omega$.

\setlength{\arraycolsep}{0.0em}
\begin{eqnarray}
Z(\Delta \omega) = \frac{R_L}{1 - jQ_L(\frac{\omega_0}{\omega_0 + \Delta \omega} -  \frac{\omega_0 + \Delta \omega}{\omega_0})}  = \nonumber\\
= \frac{R_L}{1 + jQ_L(\frac{\Delta \omega}{\omega_0} \frac{2\omega_0 + \Delta \omega}{\omega_0 + \Delta \omega})}
\end{eqnarray}
\setlength{\arraycolsep}{5pt}

Because in our aplication $\omega_0 \gg \Delta \omega$ the following assumption is possible:

\begin{equation}
 \frac{2\omega_0 + \Delta \omega}{\omega_0 + \Delta \omega} \approx 2
\end{equation}

Based on this we obtain:

\begin{equation}
Z(\Delta \omega) = \frac{R_L}{1 + \frac{2jQ_L \Delta \omega}{\omega_0}} 
\end{equation}

System with such transmission can be implemented as a digital filter with sampling period T. The bilinear transformation is used to calculate it's parameters.

\begin{equation}
j\Delta \omega = \frac{2(z-1)}{T(z+1)}
\end{equation}

\begin{equation}
Z(z) = \frac{R_L}{1+\frac{4Q_L(z-1)}{\omega_0 T(z+1)}} = \frac{R_L + R_L z^{-1}}{(\frac{4Q_L}{\omega_0T} + 1)+(1-\frac{4Q_L}{\omega_0 T})z^{-1}}
\end{equation}

\begin{figure}[!htb]
  \centering
   \includegraphics*[width=70mm]{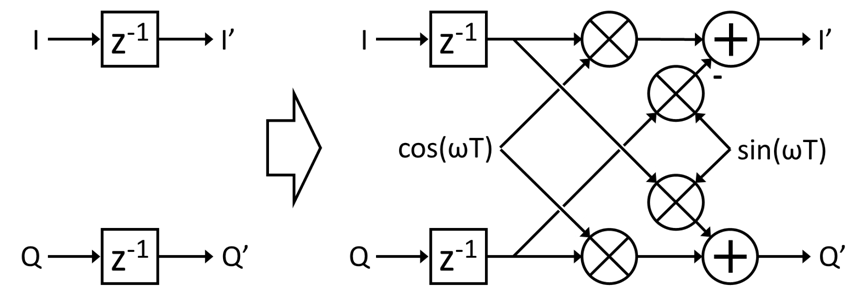}
   \caption{Change of the digital filter structure allowing the detuning.}
   \label{detun}
\end{figure}

These calculations assumed, that the cavity resonance matches precisely the reference signal frequency, but to simulate the phenomena such as piezo compensation, or Lorentz force detuning a possibility to change the resonance frequency is required. For the digital filters operating on the complex signals, it can be realized by modifying the delay block of the firmware in the way shown in Fig. \ref{detun} \cite{Complex}. To calculate the trigonometric functions Taylor's theorem is used. It requires more resources than the typically used CORDIC algorithm but needs much fewer clock cycles to obtain the result (for a 32-bit resolution it is 5 vs. 32). 

The other $\pi$-modes are simulated using filters with identical structure (but different parameters) connected in parallel. The cavity probe signal is calculated as a sum of outputs of those filters. It is assumed that all modes will detune by the same value. The detuning of the cavity is a sum of 5 components: 
\begin{itemize}
\item Lorentz Force detuning - calculated as a square of the probe signal's magnitude multiplied by a constant coefficient.
\item Piezo compensation - a sum of data from two ADCs sampling the piezo signals.
\item Microphonics  - signal generated from a look-up table.
\item Constant detuning.
\item Mode frequency - different value for each $\pi$-mode.
\end{itemize}

The detuning signal is filtered by an IIR digital filter that represents the mechanical response of the cavity. The exact transfer function of this filter will be based on the measurements performed on the real cavities. 

The beam current is calculated from the IQ samples of the RF reference signal multiplied by the data from a look-up table allowing to simulate different beam profiles. The beam current signal is added to cavity forward signal, which drives the cavity model.

\subsection{Amplifier Model}
The amplifier model is designed to simulate following phenomena:
\begin{itemize}
\item limited bandwidth,
\item gain compression,
\item power supply ripple.
\end{itemize}

The limited bandwidth is simulated with the same filter as the one used for the cavity model. The main difference is much lower Q-factor and no detuning. 
The transfer characteristic of the amplifier is stored in two lookup tables representing the gain and phase shift. These tables are addressed by the input's signal magnitude, which is estimated using "$\alpha$Max + $\beta$Min" algorithm \cite{MINMAX}. 

The power supply ripple signal is generated as a predefined waveform. Its impact on the output signal amplitude and phase can be expressed using following relations \cite{Hara, Zeng}: 

\begin{equation}
\Delta \theta= \frac{2\pi L}{\sqrt{\frac{2eV_k}{m}}}\frac{\Delta V_k}{V_k}
\end{equation}

\begin{equation}
\frac{\Delta V_{out}}{V_{out}}= \frac{5}{4}\frac{\Delta V_k}{V_k}
\end{equation}

Where $\Delta \theta$ is the output phase change, $L$ is the distance between the input and output cavities, $e$ is electron charge, $m$ is electron mass, $V_k$ is the cathode voltage, $V_{out}$ is the amplitude of the output signal. 

\subsection{Circulator Model}

The circulator is modeled as a fully linear element. It generates the cavity forward, and the amplifier reflected signal. Both of those are calculated as a sum of the cavity reflected and amplifier forward signals multiplied by a constant predefined two by two matrixes.

\section{Cavity Simulator Measurements}

To prove the proper operation of the Cavity Simulator a series of measurements was performed. The parameters of the simulations were not based on the measurements of the real cavities, because they are not yet available. Instead the values that may expose possible problems were selected. 

All measurements were performed using 704.51 MHz reference signal generated by the Cavity Simulator. The results are presented below.

\subsection{Transmission}

\begin{figure}[!htb]
  \centering
   \includegraphics*[width=70mm]{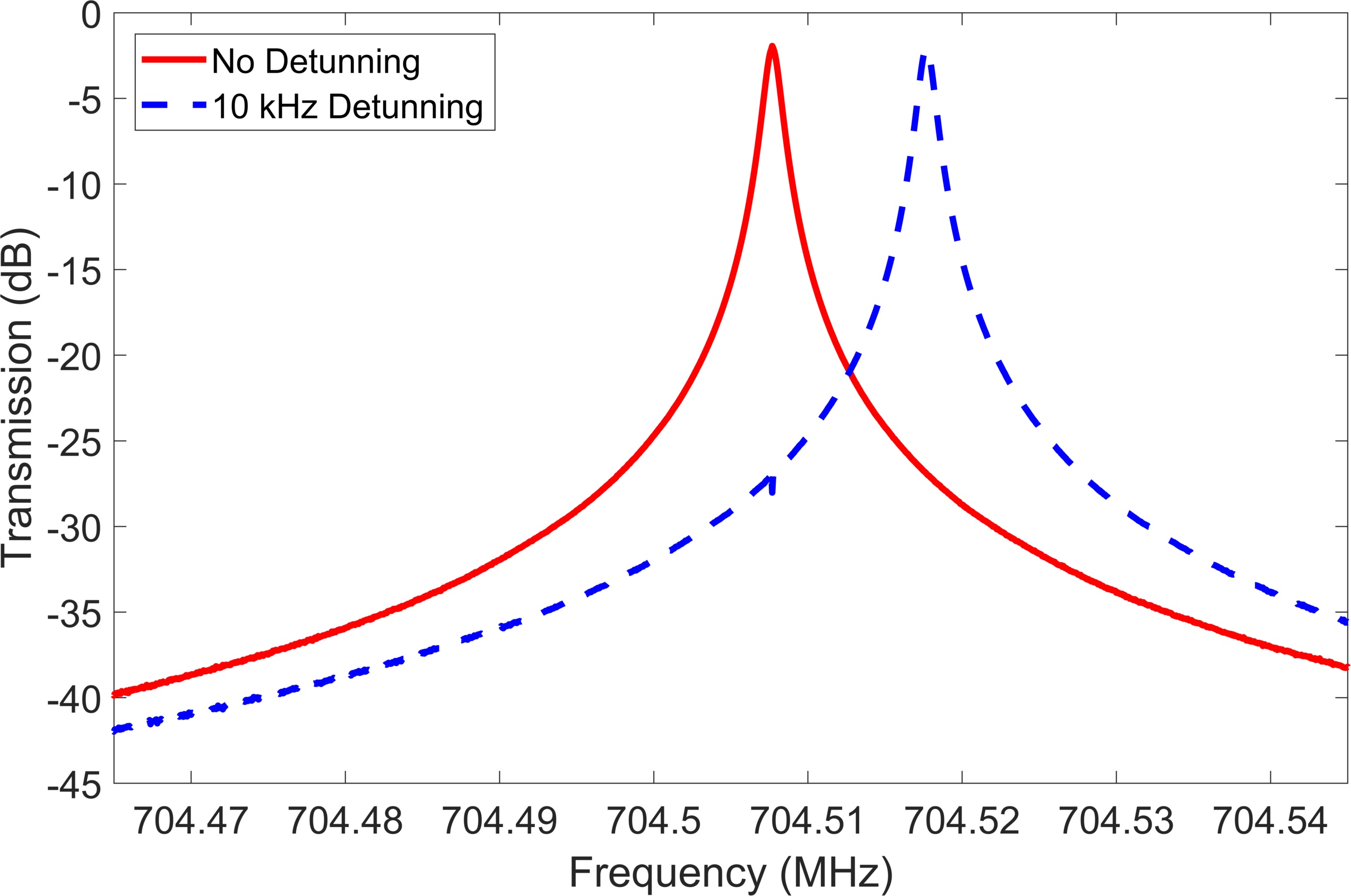}
   \caption{Transmision of Cavity Simulator with 0 and 10 kHz detunning.}
   \label{trans}
\end{figure}

\begin{figure}[!htb]
  \centering
   \includegraphics*[width=70mm]{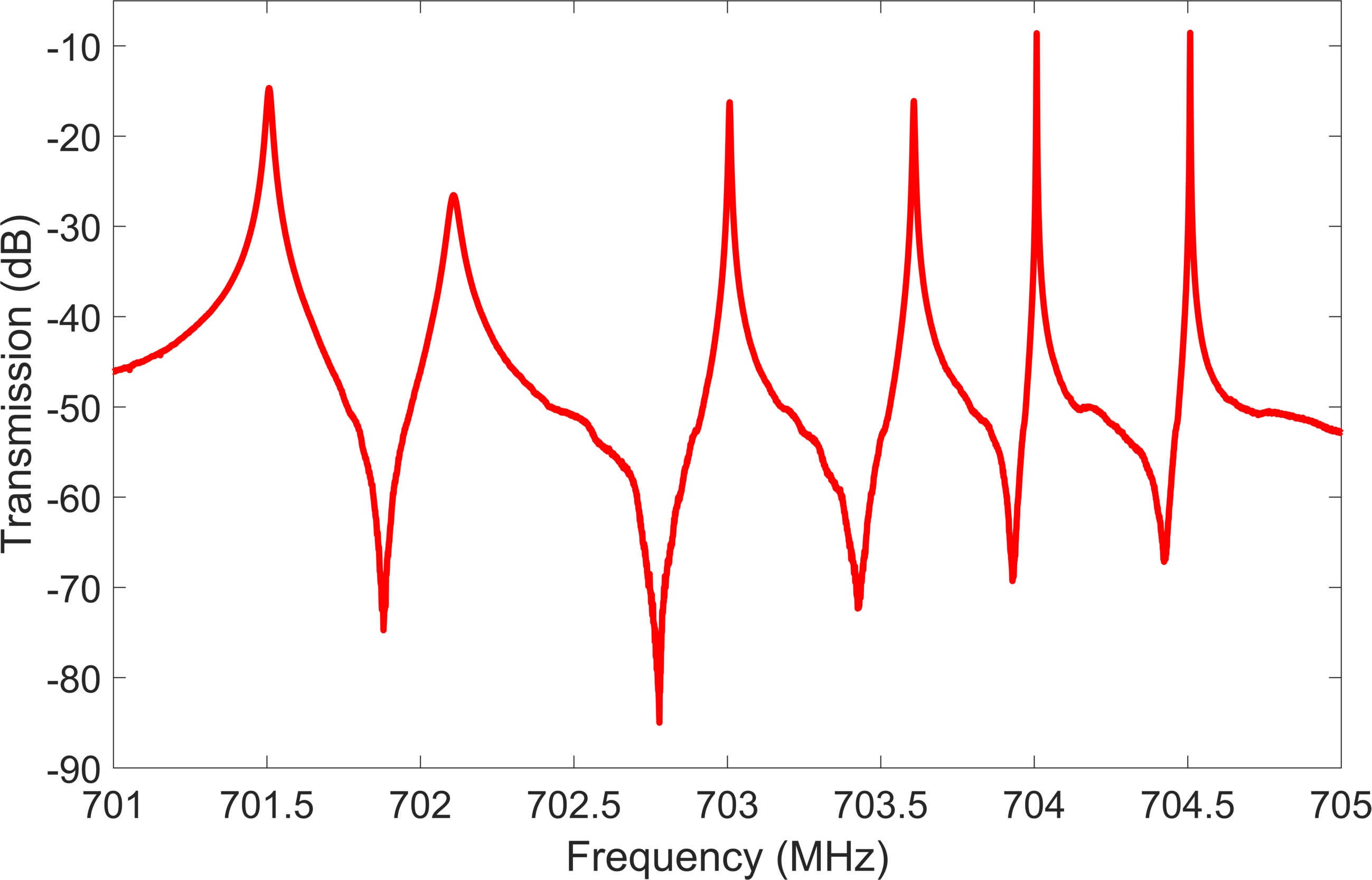}
   \caption{Transmision of Cavity Simulator with multiple $\pi$-modes }
   \label{multimode}
\end{figure}

The Cavity Simulator's transmission between the RF input and cavity probe output was measured using a vector network analyzer. The Q factor of the main mode was set to 700 000. Three measurements were performed: 
\begin{itemize}
\item one mode no detuning,
\item one mode with 10 kHz detuning,
\item 6 modes with 0, -0.5, -0.9, -1.5, -2.4, and -3 MHz offsets, all with different R and $Q_L$ settings.
\end{itemize}
Fig. \ref{trans} and \ref{multimode} present the results. 

The loaded Q-factor and maximum of the transmission don't change when the detunning is applied. Negligible distortion at the reference frequency is visible in the detunned transmission. It is caused by the limited carrier suppression of the vector modulator circuit. It can be improved by calibration of the vector modulator's IQ offsets.

\subsection{Group Delay}

As mentioned in the Section \ref{Firmware} the processing time is one of design's critical parameters. To verify it a group delay of Cavity Simulator was measured. The simulation parameters were identical with the transmission measurement with one mode. 

\begin{figure}[!htb]
  \centering
   \includegraphics*[width=70mm]{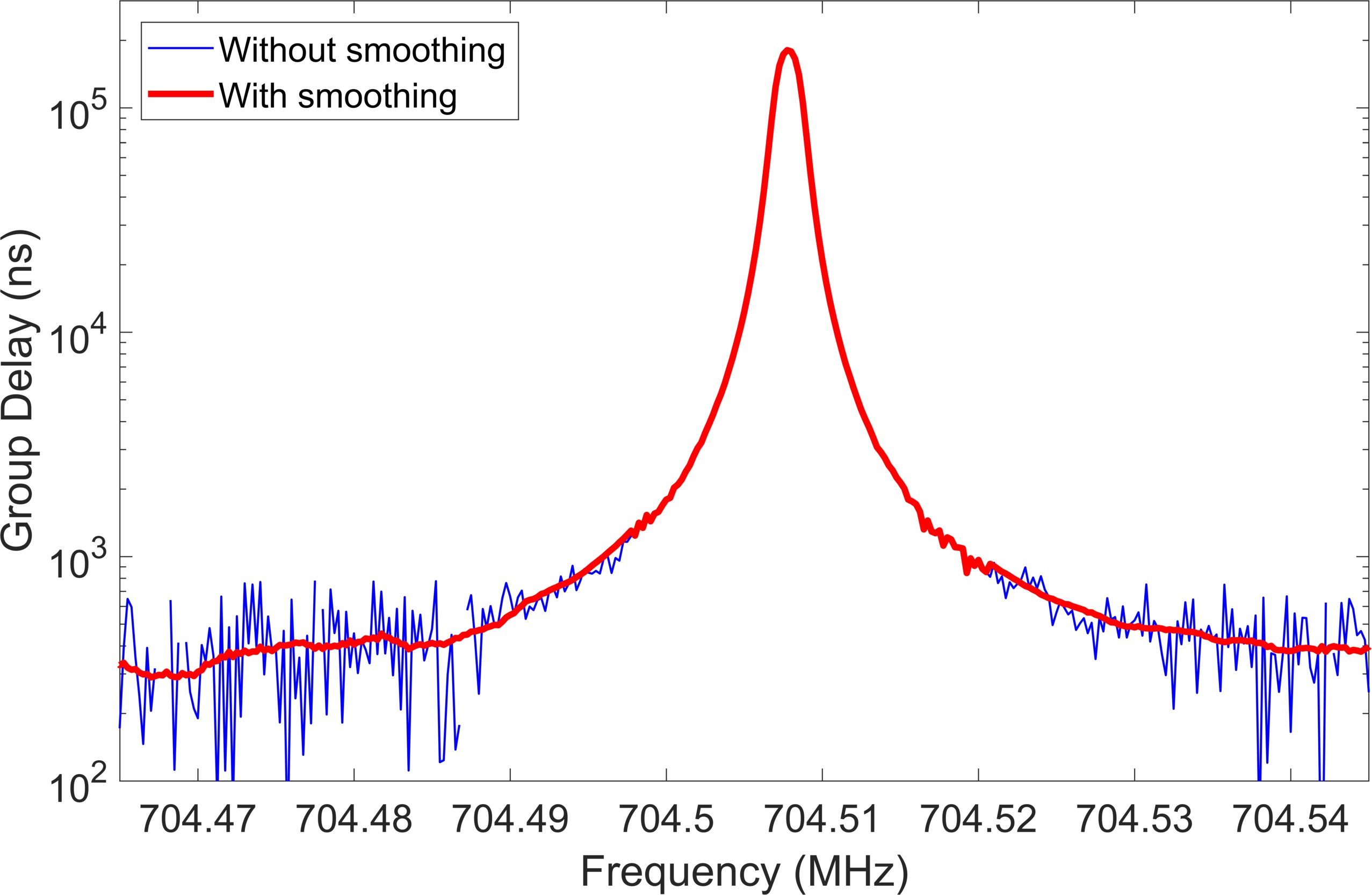}
   \caption{Cavity Simulator's group delay.}
   \label{Delay}
\end{figure}

Due to the low power of the signal at far from resonance frequencies, the results were disturbed, so smoothing of the results was applied. 
The results with and without smoothing are shown in Fig. \ref{Delay}.

The results at the frequencies close to the resonance are mostly dominated by the group delay of the cavity filter. Conversely, the values measured (after smoothing) at far from resonance frequencies indicate the processing time which is below the required 400 ns.   

\subsection{Cavity Filling and Decay}

To confirm the proper operation of the Cavity Simulator also in the time domain, the cavity filling and decay was measured. The Cavity Simulator's output was driven with RF signal generated by the 7th RF output channel. The signal was turned on and off with 2 s period.

\begin{figure}[!htb]
  \centering
   \includegraphics*[width=70mm]{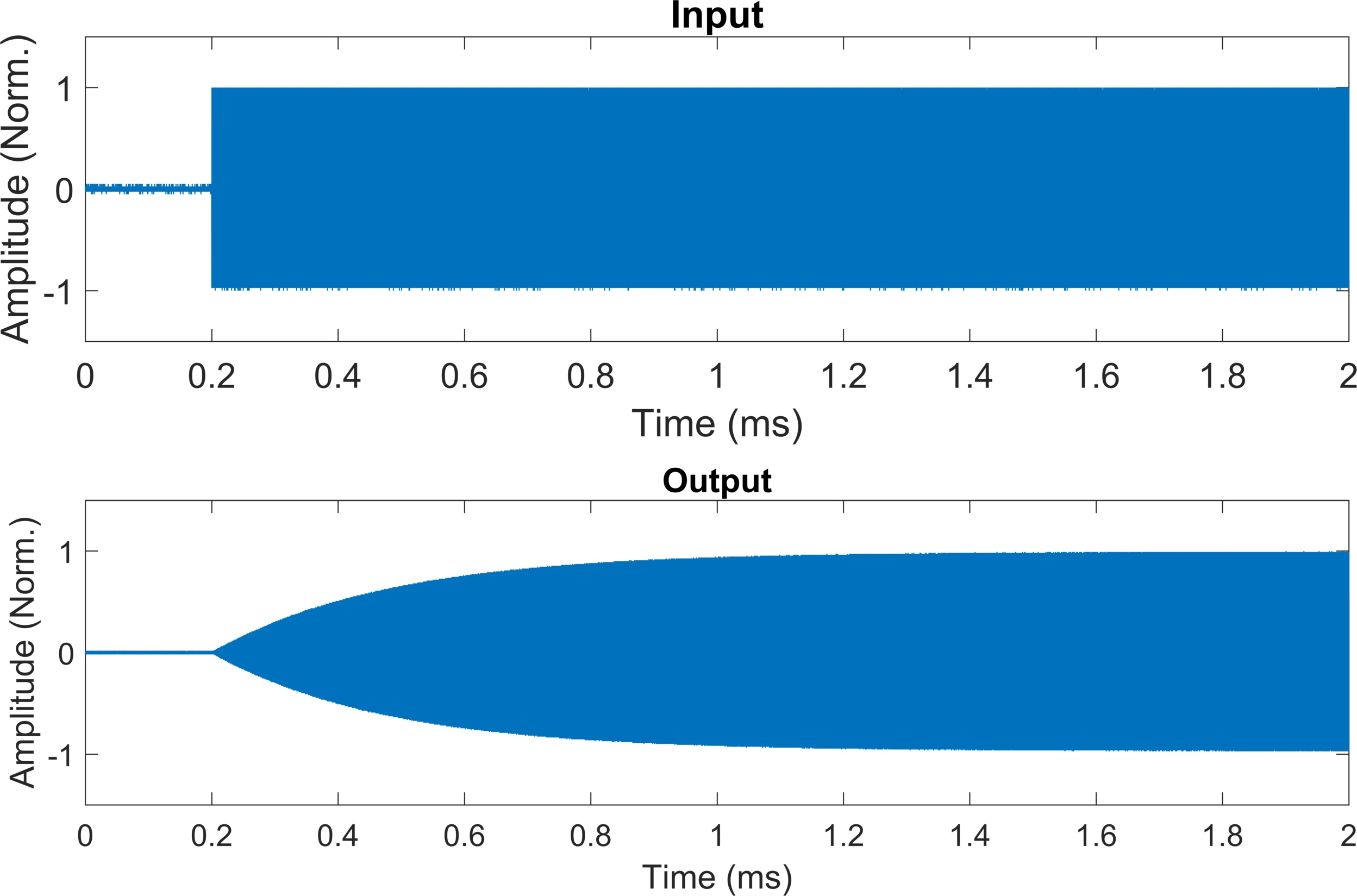}
   \caption{The input and output signals during cavity filling.}
   \label{filing}
\end{figure}

\begin{figure}[!htb]
  \centering
   \includegraphics*[width=70mm]{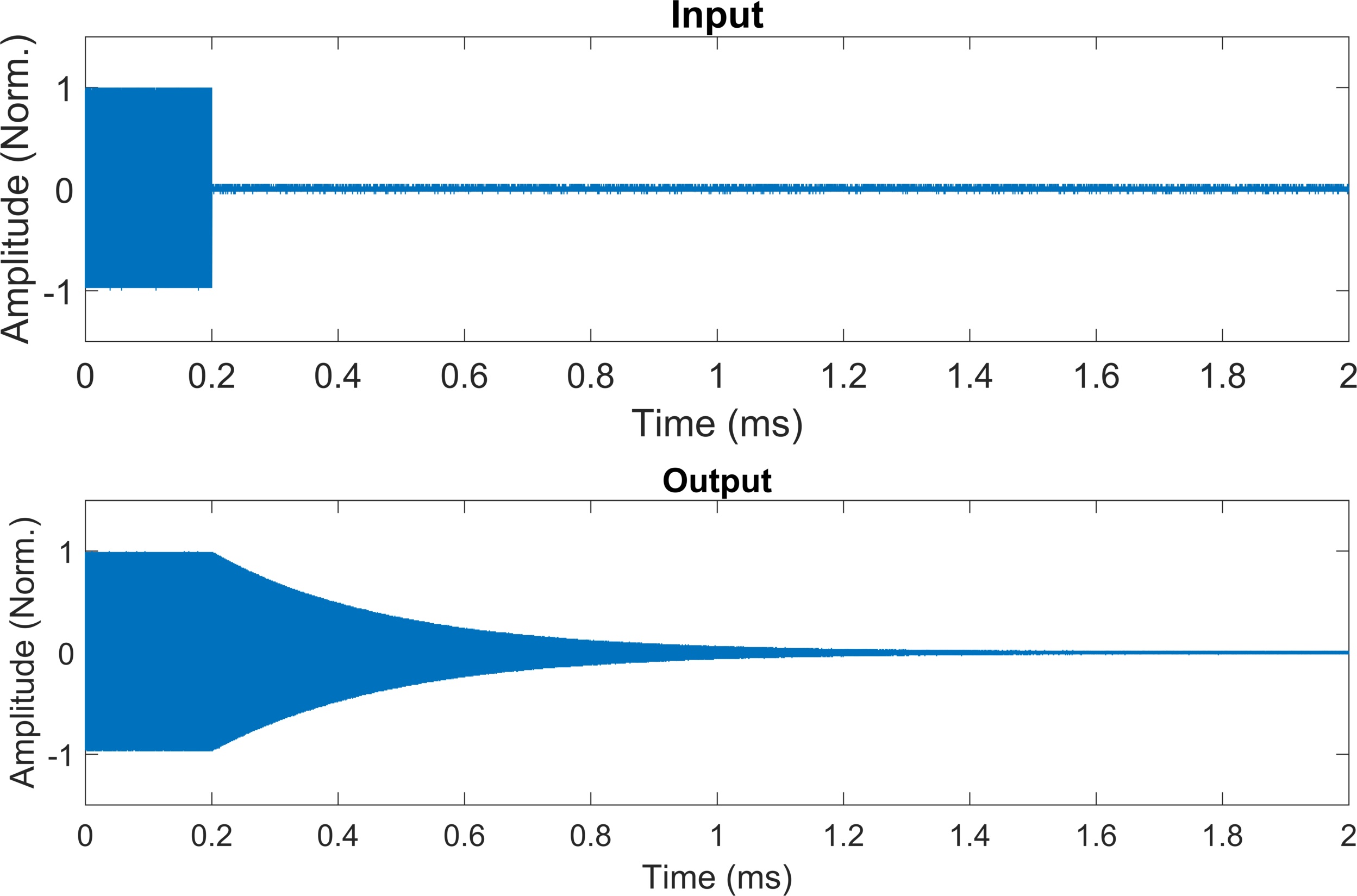}
   \caption{The input and output signals during cavity decay.}
   \label{decay}
\end{figure}

The cavity filling and decay were recorded using fast digital oscilloscope. The results are presented in Fig. \ref{filing} and \ref{decay}. The settings were identical to those used in the transmission measurement.

The filling and decay times are approximately equal to 0.65 ms, which is within the expected range. No distortions of the signals were observed.

\section{Conclusion}
In this paper, the hardware design, firmware, and measurements of the Cavity Simulator designed for European Spallation Source are described. Four units were manufactured, and the firmware with the main functionality was prepared. The measurements results confirm the proper operation of the device.

The Cavity Simulator is being further developed to extend its functionality and scope of the simulation. Next tests, including those with LLRF control System, are also planned.

\section*{Acknowledgment}

Work supported by Polish Ministry of Science and Higher Education, decision number DIR/WK/2016/03.

\ifCLASSOPTIONcaptionsoff
  \newpage
\fi



\bibliographystyle{IEEEtran}

\begin{thebibliography}{14}


\bibitem{linac} M.~Eshraqi et al.,  \emph{The ESS Linac}, IPAC 2014, Dresden, Germany, 2014.
\bibitem{LLRF}	A.~J.~Johansson, A.~Svensson, F.~Kristensen and R.~Zeng,  \emph{LLRF System for the ESS Proton Accelerator}, IPAC 2014 Dresden, Germany, 2014.
\bibitem{PEG} J.~Szewiński et al.,  \emph{Contribution to the ESS LLRF System by Polish Electronics Group}, IPAC 2017, Copenhagen, Denmark, 2017.
\bibitem{cavity} G.~Devanz et al.,  \emph{ESS Eliptical Cavities and Cryomodules}, SRF 2013, Paris, France, 2013.
\bibitem{DESY}  T.~Czarski, K.~T.~Pozniak, R.~S.~Romaniuk, and S.~Simrock,  \emph{TESLA cavity modeling and digital implementation in FPGA technology for control system development}, Nuclear Instruments and Methods in Physics Research Section A: Accelerators, Spectrometers, Detectors and Associated Equipment, Volume: 548, 2005.
\bibitem{LBNL}  C.~Serrano, L.~Doolittle and V.~K.~Vytla,  \emph{Cryomodule-On-Chip Simulation Engine}, ICALEPCS 2017, Barcelona, Spain, 2017.
\bibitem{KEK} F.~Qiu et al.,  \emph{Real-time cavity simulator-based low-level radio-frequency test bench and applications for accelerators}, Physical Review Accelerators and Beams, Volume: 21, Issue: 3, 2018.
\bibitem{LO} I.~Rutkowski, K.~Czuba and M.~Grzegrzółka,  \emph{LO Board For 704.42 MHz Cavity Simulator For ESS},  ICALEPCS 2017, Barcelona, Spain, 2017.
\bibitem{noniq} L.~Doolittle, H.~Ma and M.~S.~Champion,   \emph{Digital Low-Level RF Control Using Non-IQ Ssampling} LINAC 2006, Tennessee, USA, 2006.
\bibitem{Schilcher}  T.~Schilcher, \emph{Vector Sum Control of Pulsed Accelerating Fields in Lorentz Force Detuned Superconducting Cavities}, Ph.D. thesis, Universität Hamburg, 1998.
\bibitem{Complex} K.~W.~Martin, \emph{Complex Signal Processing Is Not Complex}, IEEE Transactions on Circuits and Systems I: Regular Papers, Volume: 51, Issue: 9, 2004.
\bibitem{MINMAX} Richard~G.~Lyons, \emph{Understanding Digital Signal Processing}, Prentice Hall PTR, 2004.
\bibitem{Hara} M.~Hara, T.~Nakamura and T.~Ohshima, \emph{A Ripple Effect Of a Klystron Power Supply On Synchrotron Oscillation}, Particle Accelerators, Volume: 59, 1998.
\bibitem{Zeng} R.~Zeng, D.~McGinnis, S.~Molloy and A.~J.~Johansson, \emph{Influence of the Droop and Ripple of Modulator on Klystron Output},  IPAC2012, New Orleans, Louisiana, USA, 2012.


\end{thebibliography}
\end{document}